\documentclass[psamsfonts]{amsart}
\usepackage{amsmath}
\usepackage{amsfonts,amssymb}
\usepackage{graphicx}
\usepackage[all,knot,poly]{xy}

\newtheorem{thm}{Theorem}[]

\begin{document}

\title[The Fundamental Group of a Spatial Section]{The Fundamental Group of a Spatial Section Represented by a Topspin Network}

\author{Christopher L Duston}
\address{Stony Brook University, Stony Brook, New York, 11974}
\email{christopher.duston@stonybrook.edu}



\begin{abstract}
We present an algorithm which determines the fundamental group of a spatial section using topspin networks. Tracking the topology of the spatial section is a unique feature of this approach, which is not possible in standard Loop Quantum Gravity. This leads to an example of spatial topology change in a smooth 4-manifold represented by a topspin foam.
\end{abstract}

\maketitle

\section{Topology in Loop Quantum Gravity}
The phenomena of topology change in classical gravity is generally accepted to be restricted to changes which preserve causality\cite{Anderson-DeWitt}. In the quantum case, very little is known about topology change but it certainty depends on the specific model being considered. For instance, one can study instatons in semiclassical Euclidean quantum gravity\cite{Gibbons-2011}, but the underlying quantum structure is not known. Discrete approaches to quantum gravity (such as causal sets or dynamical triangulations\cite{Ambjorn-Budd-2013}) naturally include some topological information, but there is not much connection to the classical theory.

Currently, the best candidate for a quantum theory of gravity is loop quantum gravity (LQG). In short, it is a quantization of the gravitational field following the canonical approach of Dirac (for a review, see the excellent texts \cite{Rovelli, TTbook}). It is well-defined from the point of view of mathematical physics, and may shortly be open to experimental scrutiny \cite{Barrau-etal-2012}. The key feature of LQG which will be relevant for us is the role of spin networks - these are graphs embedded in spatial sections with the fixed topology of $\mathbb{S}^3$. The edges of the graphs are decorated with irreducible representations $\rho_i$ of $SU(2)$, and the vertices are labeled with intertwiners that map $n$ ingoing to $m$ outgoing representations:
\[i_v:\rho_1\otimes \rho_2\otimes...\otimes\rho_n \to \tilde{\rho}_1\otimes \tilde{\rho}_2\otimes...\otimes\tilde{\rho}_m.\] 
In the case that these intertwiners are invariant subspaces, the spin networks represent gauge-invariant states and are natural generalizations of Wilson loops. 
Spin networks provide information about the geometry of the spatial section via the $SU(2)$ spin labels, with higher spin generally corresponding to larger curvature. However, \textit{spin networks are ignorant of the topological structure of the spatial section}. It is easy to see why; by restricting from $\mathbb{S}^3$ to the discrete information on the graph, the resulting spin network could be embedded in any number of topologically inequivalent 3-manifolds. The exact dynamics of LQG has still not been settled, but a dynamical change to a spin network requires one of two \textit{ad hoc} specifications:
\begin{itemize}
\item Assume the topology is unchanged from the trivial one ($\mathbb{S}^3$), or
\item Disregard all topological information about the spatial section that the new spin network is embedded in.
\end{itemize}
The first choice forces the topology to remain background-dependent under quantization, and the second choice leads to serious problems when trying to take the classical limit of the theory (essentially forcing an arbitrary specification of the topology). The true role of background independence in fundamental theories is not know (see \cite{Smolin-2005} for a careful discussion), but it is reasonable to expect that a quantized theory of gravity should help us to understand the topological nature of the universe. Thus, neither of the choices given above should be considered satisfactory, and this problem serves as the major motivation for this work.

\section{Topspin Networks}
We seek a solution to the loss of topological data in LQG which will also preserve the existing geometric structure provided by the spin networks. The solution we present is based on the following classic theorem of Alexander\cite{Alexander-1920}:

\begin{thm}
Any compact oriented 3-manifold can be described as a branched covering of $\mathbb{S}^3$, branched along a graph $\Gamma$.
\end{thm}

The topological structure of the covering spaces is inherited from the base, except over the branch locus where closed curves can travel into multiple components of $\mathbb{S}^3\setminus \Gamma$. To characterize how this happens, each edge of the graph over which an $n$-fold cover is branched can be labeled with an element of $S_n$. In other words, we can specify a representation $\sigma:\pi_1(\mathbb{S}^3\setminus \Gamma)\to S_n$, which allows us to recover an arbitrary smooth 3-manifold by a specification of the graph $\Gamma$ and a set of topological labels $\{\sigma_i\}$ for each edge $e_i$. 

Alexander's theorem is actually quite powerful, as it can be used to represent $n>2$ dimensional manifolds branched over $n-2$ complexes in $n$-spheres. In the case of dimension 4, it can even be specialized further to branch loci which are embedded surfaces\cite{Iori-Piergallini-2002}. Using an equivalent specification as above, this method can be used to represent both the geometric and topological data of the gravitational field in terms of an embedded surface $\Sigma$ and the representation $\sigma:\pi_1(\mathbb{S}^4\setminus \Sigma)\to S_n$. This approach is explored further in \cite{Duston-2013}.

\begin{figure}\begin{center}
\includegraphics[scale=0.4]{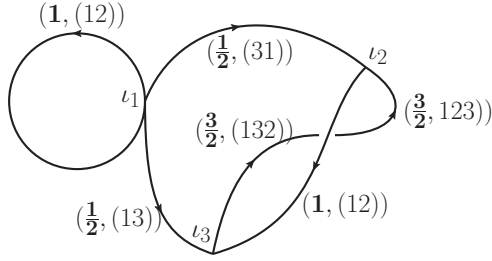}
\caption{An example of a topspin network.}\label{fig:Topspin_Ex}
\end{center}\end{figure}

The proposed connection with LQG is through the graphs; in the topspin formalism, the spin network graphs are identified with the branch locus and the representation $\sigma$ is added through a labeling of each edge with an element of $S_n$ (``topological labels'')\cite{Marcolli-alYasry-2008,DMA}. An example of a topspin network is given in figure \ref{fig:Topspin_Ex}. The two sets of labels can always be made compatible by adding trivial representations to the branch locus and the trivial element $(1)\in S_n$ to the spin network. Some modifications to the algebra of LQG is required to make the holonomies well-defined; specifically, the fields now take values in the algebra $\mathcal{U}(\mathfrak{su}(2))\otimes \mathbb{C}S_n$ rather than $\mathfrak{su}(2)$. This leads to some slight redefinitions of the states and operators, but since this approach exploits natural degeneracies in the spin networks, the modifications are largely notational. For more details on these modifications see \cite{Duston-2012}.

The representation of a specific branched covering space is not unique; there is a set of covering moves for these graphs, analogous to the Reidemeister moves but which are consistent with both the spin and topological labels \cite{DMA}. This allows one to determine when two topspin networks represent topologically equivalent spatial sections. The purpose of the current work is to calculate the fundamental group of a given spatial section, and present a topspin foam which represents a topology-changing amplitude for the quantum gravitational field.

\section{The Fox Algorithm and Examples}
\begin{figure}
\begin{center}\leavevmode
\begin{xy}
(0,0)*+{\Sigma\setminus p^{-1}(\Gamma)}="S-G"; (0,-15)*+{\mathbb{S}^3\setminus\Gamma}="S3-G";(20,-15)*+{\mathbb{S}^3}="S3";(20,0)*+{\Sigma}="S"; (12,0)*+{\subset};(12,-15)*+{\subset};
{\ar@{->} "S-G";"S3-G"}?*!/_2mm/{};
{\ar@{->} "S";"S3"}?*!/_2mm/{p};
\end{xy}
\caption{The setup for the modified Fox algorithm. The spatial section $\Sigma$ is a branched cover of $\mathbb{S}^3$, branched over the graph $\Gamma$.}\label{fig:branched_cover}
\end{center}
\end{figure}
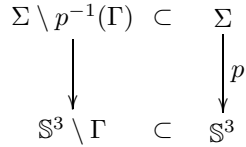

The method which we present to determine the fundamental group of a spatial section represented by a topspin network is based on the equivalent problem in knot theory, first presented by Fox \cite{Fox-1962} and based on a classic theorem of Reidemeister and Schreier \cite{Combinatorial_Group_Theory}. The setup for this algorithm is illustrated in figure \ref{fig:branched_cover}. For the moment we assume the cover is connected. The disconnected case is particularly important for topological change - for instance, the merging of two spatial spheres - but we will save this generalization for a later work. 

\begin{figure}\begin{center}
\includegraphics[scale=0.25]{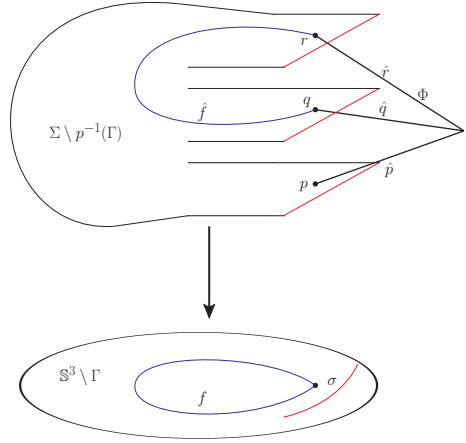}
\caption{Attaching a frame $\Phi$ to a connected cover of $\mathbb{S}^3$.}\label{fig:frame}
\end{center}\end{figure}

First we present the algorithm and then proceed to show a few examples.
\begin{itemize}
\item Find a presentation $\langle x_1,...,x_n|r_1,...,r_m\rangle$ of the graph group $G=\pi_1(\mathbb{S}^3\setminus\Gamma)$. This is given by a generator $x_i$ on each edge and a set of relations $r_j$ on each crossing. For the case of graphs, we also need to include the vertex relation
\[x_1...x_px^{-1}_{p+1}...x^{-1}_{p+q}=1\]
for $p$ ingoing edges and $q$ outgoing edges.
\item Attach a $g$-frame $\Phi$ to the covering space as shown in figure \ref{fig:frame}. Closed curves in $\mathbb{S}^3\setminus\Gamma$ are now represented in $(\Sigma\setminus p^{-1}(\Gamma))\cup\Phi$ by paths $\hat{p}\hat{r}^{-1}\hat{f}\hat{q}\hat{p}^{-1}$, and the fundamental group of that space is $H\ast F_{g-1}$, where $F_{g-1}$ is a free group which is completely due to the addition of the frame.
\item There is a homomorphism\cite{Fox-1962} which maps any word in $G$ to one in $H\ast F_{g-1}$: 
\[u=x_{j_1}^{\epsilon_1}x_{j_2}^{\epsilon_2}...\to u_{\alpha}=x_{j_1\alpha_1}^{\epsilon_1}x_{j_2\alpha_2}^{\epsilon_2}...,\]
where $\alpha_k=\sigma_k(\alpha)$ and the permutation elements $\sigma_k$ are given by the following rule:
\[\epsilon_k=\left \{\begin{array}{cc}
+1,&\sigma_k=\sigma(x_{j_{k-1}}^{\epsilon_{k-1}})...\sigma(x_{1}^{\epsilon_1})\\
-1,&\sigma_k=\sigma(x_{j_{k}}^{\epsilon_{k}})...\sigma(x_{1}^{\epsilon_1})\end{array}\right.\]
Here the $\sigma(x)$ are the topological labels on each edge.
\item Now the free group must be removed. This is done by identifying a Schreier tree (a set of words $w_i$ such that the left segment of each word is one of the other words) and adjoining the relations which trivialize their images in the cover:
\[w_{10}=w_{11}=...=w_{g-1,0}=1.\]
This mimics the original Reidemeister-Schreier algorithm since each one of these words is associated to a different sheet of the cover. This gives us the fundamental group $H=\pi_1(\Sigma\setminus p^{-1}(\Gamma))$.
\item The images of the branch locus are elements of the kernel of the homomorphism
\[\pi_1(\Sigma\setminus p^{-1}(\Gamma))\to \pi_1(\Sigma),\]
so by adjoining the relations which trivialize them we can recover $\pi_1(\Sigma)$ by the first isomorphism theorem. For such an element $v$ of the branch locus represented by $\rho(v)=(\beta_1...\beta_\lambda)(...)...$ (in cycle notation) we adjoin the relations
\[v_{\beta_1}...v_{\beta_\lambda}=1,...\]
\end{itemize}

We now illustrate with some examples.

\begin{figure}
\begin{minipage}{0.45\textwidth}
\begin{center}
\includegraphics[scale=0.3]{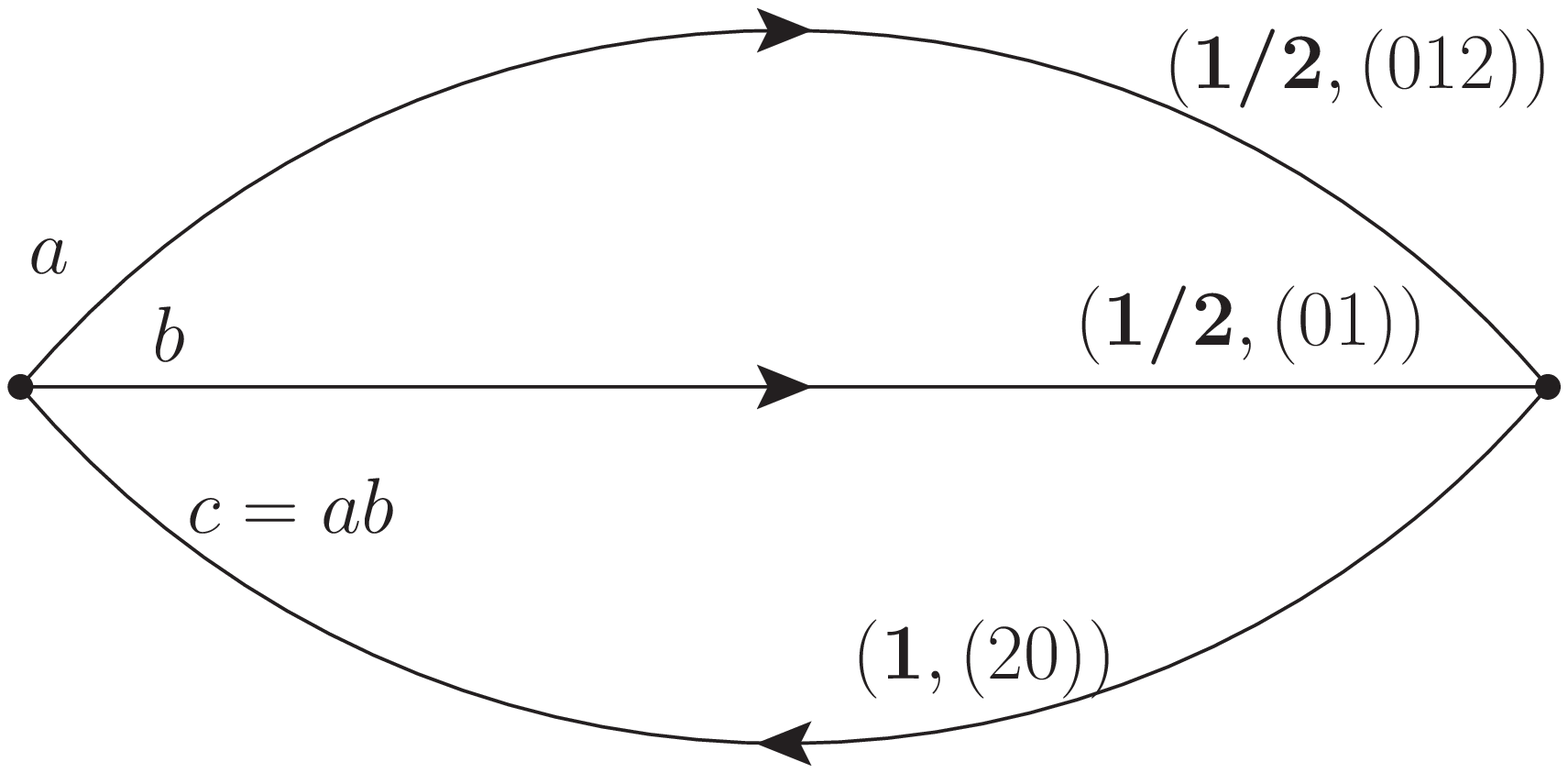}
\caption{2-vertex topspin network representing $\Sigma_2$.}\label{fig:Simple_Topspin}
\end{center}
\end{minipage}
\begin{minipage}{0.45\textwidth}
\begin{center}
\includegraphics[scale=0.2]{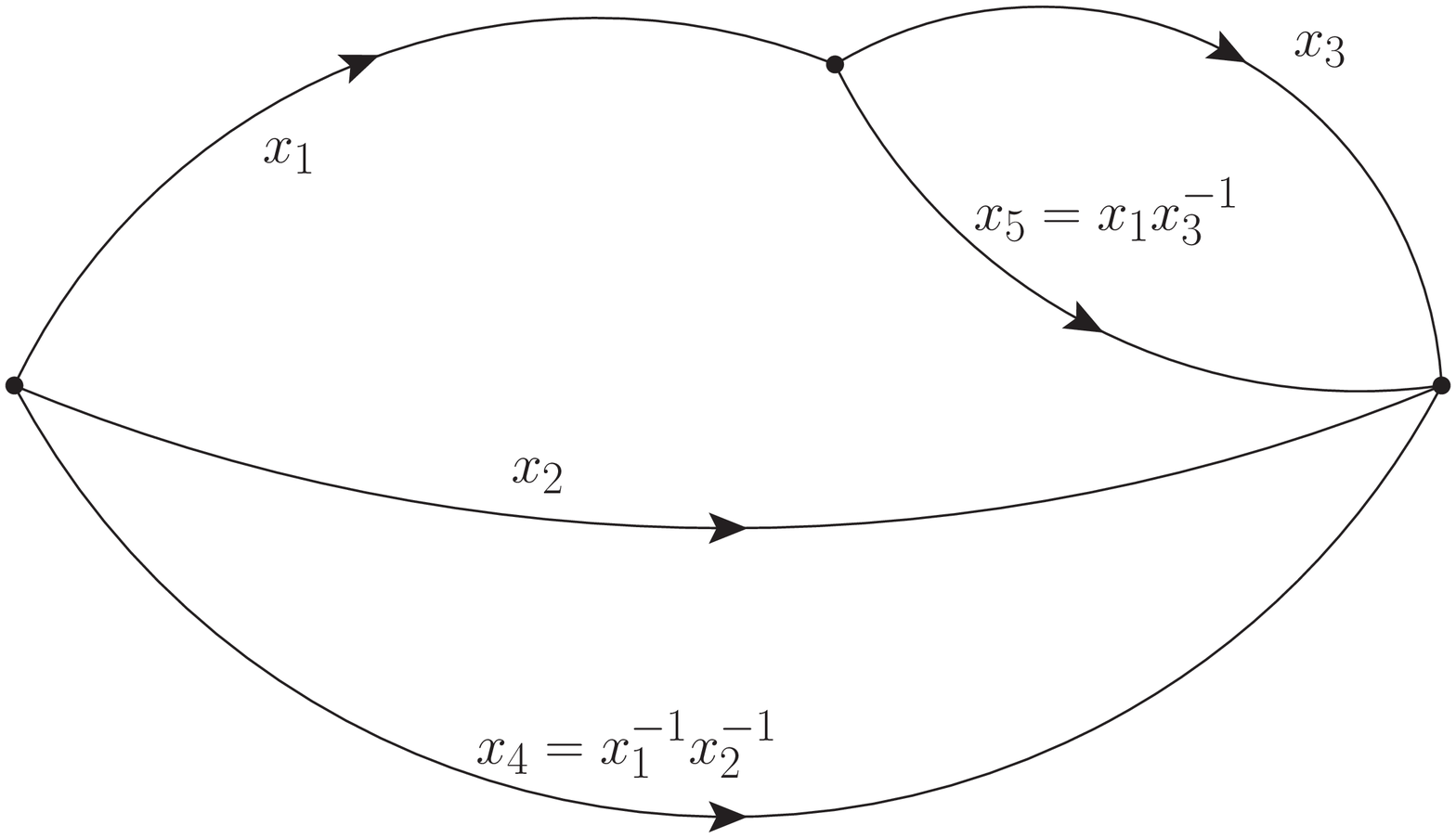}
\caption{3-vertex topspin network representing $\Sigma_4$.}\label{fig:Simple_Topspin_2}
\end{center}
\end{minipage}
\end{figure}

\medskip
\noindent\textit{Example 1:} 2 vertices, 3 edges (see figure \ref{fig:Simple_Topspin}). The edges are generated by $a,b,c$ and the third edge is restricted by the vertex (the relations from the two vertices are duplicate). The group is thus $G=\langle a,b,c|c=b^{-1}a^{-1}\rangle$. We will use the representation $\sigma(a)=(012)$ and $\sigma(b)=(01)$ (forcing $\sigma(c)=(20)$). Under the homomorphism described above we have
\[H \ast F_2=\langle a_0,a_1,a_2,b_0,b_1,b_2,c_0,c_1,c_2|c_0=a_0b_1,~c_1=a_1b_2,c_2=a_2b_0 \rangle.\]
To remove the free group we choose $w_0=1$, $w_1=a$, and $w_2=a^2$ (which satisfies the Schreier condition). To find the image of these words in $H\ast F_2$ we can use the following tables (and with $w_{0,\alpha}=1~\forall\alpha$):

\begin{equation*}w_1:\begin{array}{|c|cc|c}
\alpha&\sigma_1&\alpha_1&w_{1\alpha}\\
\hline
0&(1)&0&a_0\\
1&(1)&1&a_1\\
2&(1)&2&a_2\end{array}
\qquad w_2:\begin{array}{|c|cccc|c}
\alpha&\sigma_1&\alpha_1&\sigma_2&\alpha_2&w_{2\alpha}\\
\hline
0&(1)&0&(012)&1&a_0a_1\\
1&(1)&1&(012)&0&a_1a_0\\
2&(1)&2&(012)&2&a_2a_1\end{array}\end{equation*}
Now by setting $w_{j0}=1$ we get $a_0=1$ and $a_0a_1=1\rightarrow a_1=1$. The fundamental group of the sheet complement is thus reduced in the following way
\begin{align*}
H&=\langle a_0,a_1,a_2,b_0,b_1,b_2,c_0,c_1,c_2 | c_0=a_0b_1, c_1=a_1b_2,c_2=a_2b_0\rangle\\
&=\langle a_2,b_0,b_1,b_2,c_0,c_1,c_2 | c_0=b_1, c_1=b_2, c_2=a_2b_0\rangle\\
&= \langle a_2, b_0, b_1, b_2\rangle .
\end{align*} 
The kernel of $ H \to\pi_1(\Sigma)$ is
\[a_0a_1a_2=1,~b_0b_1=b_2=1,~c_2c_0=c_1=1,\]
and since the last relation tells us $(ab)_2(ab)_0=a_2b_0a_0b_1=1$, we find that all of the generators are trivial and so finally
\[\pi_1(\Sigma_2)=0.\]

\medskip
\textit{Example 2:} 3 vertices, 5 edges (see figure \ref{fig:Simple_Topspin_2}). This is one of the networks which would result from the action of the Hamiltonian on figure \ref{fig:Simple_Topspin}. Here the graph group is $G=\langle a,b,c,d,e,f|c=ba,e=da,f=a^{-1}d^{-1}b^{-1}\rangle$ and the various representations are given in the figure. The fundamental group of the cover complement with the frame is

\begin{equation*}
H\ast F_2=\left(\begin{array}{c|c}
a_0,b_0,c_0,d_0,e_0,f_0&c_0=b_0a_2,e_0=d_0a_1,f_0=a^{-1}_2d_1^{-1}b_2^{-1}\\
a_1,b_1,c_1,d_1,e_1,f_1&c_1=b_1a_0,e_1=d_1a_2,f_1=a^{-1}_0d_2^{-1}b_0^{-1}\\
a_2,b_2,c_2,d_2,e_2,f_2&c_2=b_2a_1,e_2=d_2a_0,f_2=a^{-1}_1d_0^{-1}b_1^{-1}
\end{array}\right)\end{equation*}

Choosing our Schreier tree to be $1, a,a^2$ we find $a_0=a_1=1$ and the fundamental group is 
\[\pi_1(\Sigma_4)=\langle  d_0,d_1,d_2|d_2=d_1^{-1}d_0^{-1}\rangle \simeq\mathbb{Z}\ast \mathbb{Z}.\]

\begin{figure}\begin{center}
\includegraphics[scale=0.3]{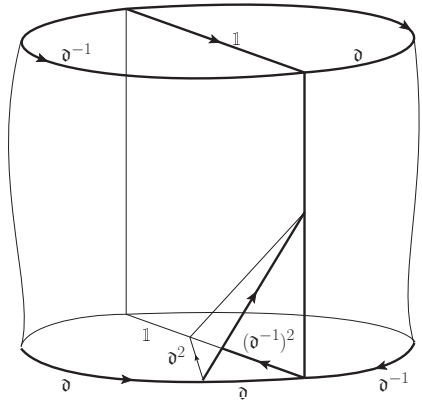}
\caption{A simple example of a topspin foam with boundaries that are topologically different spatial sections.}\label{fig:Simple_Foam}
\end{center}\end{figure}

\medskip
\textit{Example 3 (Topspin Foam):} There is an analogue of spin foams in this topological framework - topspin foams \cite{DMA}. The previous two examples easily lead to a topspin foam whose boundaries are topologically inequivalent spatial sections just by ensuring that all the edge relations are satisfied. This is shown in figure \ref{fig:Simple_Foam}. In the figure we have defined $\mathfrak{d}$ to be any cycle of maximum length in $S_3$ - \textit{i.e.} (012), (021) or inverses. It should be noted that since topspin foams which satisfy the vertex and edge conditions represent smooth 4-manifolds, this example represents a smooth transition between inequivalent spatial sections.


\medskip
Finally, there is a simple way to extend example 1 to cover all 2-vertex topspin networks:

\begin{thm}
Regardless of how many edges, a 2-vertex topspin network represents a spatial section with trivial fundamental group.
\end{thm}

The full proof of this result will be presented in a future work. This is particularly relevant for loop quantum cosmology, where such simple networks are \textit{assumed} to have the topology of a sphere \cite{Rovelli-Vidotto-2008, Borja-etal-2011}.

\section{Summary and Outlook}
In this short communication we have summarized how one can explicitly determine the fundamental group of a spatial section using the topspin network formalism. We have presented some simple examples, as well as a topspin foam which represents a cobordism between a trivial and a non-trivial spatial section. This topspin foam is generated by the action of the Hamiltonian, and is a model for topology change in LQG, unique to the topspin approach. 

Although we have not elaborated on the other topological characteristics of the spatial section which topspin networks represent, it should be noted that since they are closed, the homology groups are given by this construction as well. The first homology group is $H_1=\pi_1/[\pi_1,\pi_1]$, and under Poincar\'e duality we have $H_1\simeq H^2$. The modified Fox algorithm thus provides a complete specification of both the homotopy and homology groups which classify the spatial sections.
\medskip

\thanks{This work is based on a presentation given by the author during the LOOPS 13 conference at the Perimeter Institute in Waterloo, Canada. Travel support was provided by the FGSA Travel Award for Excellence in Graduate Research.} 

\bibliographystyle{plain}
\bibliography{ref}


\end{document}